# A POSSIBLE ORIGIN OF C-TYPE ASTEROIDS AND CARBONACEOUS CHONDRITES

## V. V. BUSAREV

Moscow Lomonosov State University, Sternberg Astronomical Institute (SAI), 119991 Moscow, Russian Federation

**Author's address:** Lunar and Planetary Dep., Sternberg State Astronomical Institute, 119991 Moscow, Universitetskij av., 13, Russian Federation; e-mail: **busarev@sai.msu.ru**

# ABSTRACT

A hypothesis based on observational and theoretical results about the origin of C-type asteroids and carbonaceous chondrites is put forward. Asteroids of C-type and close BGF-types could form from hydrated silicate-organic matter accumulated in the cores of water-differentiated bodies existed in the growth zone of Jupiter and, possibly, Saturn. The gravitational scattering of such bodies by Jupiter at its final stage of formation to the main asteroid belt might have led to fragmentation and re-accretion of their primitive matter on the surfaces of many asteroids and/or asteroid parent bodies. Similarly, asteroids of other primitive types (D and P) characterized, probably, elevated content of organics could origin from the matter of small bodies accreted and evolved in the growth zones of Uranus and Neptune and then dispersed by the giant planets to the asteroid belt. The hypothesis makes clear a row of long-standing puzzling facts, the main of which are as follows. The low-albedo and carbonaceous-chondritic surface properties of (1) Ceres contradict to its probable differentiated structure and icy crust (e. g., Thomas *et al.*, 2005, *Nature* 437: 224-226; Castillo-Rogez *et al.*, 2010, Icarus 205, 443–459), but it could be explained by the process of primitive matter fall. Atypical hydrated silicates (probably, as a component of carbonaceous-chondritic matter) are found on the surfaces of many asteroids of high-temperature types (Rivkin *et al.*, 1995, *Icarus* 117: 90-100; Rivkin *et al.*, 2000, *Icarus* 145: 351-368; Busarev, 1998, *Icarus* 131: 32-40; Busarev, 2002, *Solar System Research* 36: 39-47) that may be a consequence of prolonged precipitation of carbonaceous-chondritic matter on their surfaces. Some carbonaceous chondrites are unidirectionly magnetized (e. g., Stacey *et al.*, 1961, *Geophysical Research Journal* 66: 1523-1534; Butler, 1972, *Earth and Planetary Science Letters* 17: 120-128) that may be an indication of their stay on the surface of a differentiated body having strong magnetic field.

## INTRODUCTION

The most reliable method of solid celestial bodies study is a direct and comprehensive investigation of meteorites and samples of the bodies delivered by astronauts and spacecrafts. In this respect, it is very illustrative an example of discovery of the early thermal evolution of solid celestial bodies. Detailed analyses of chondritic and achondritic meteorites (Lee *et al.*, 1976; Srinivasan *et al.*, 1999) showed that there was a period of decay of short-lived radionuclides (mainly $^{26}$Al and others, such as $^{60}$Fe, etc. contained in silicates, e. g., Goswami, 2004) accompanied by a considerable heat release. The isotopes could be trapped in the matter of the early solar system because of the nova and super-nova explosions. It is confirmed by the discovering of $^{26}$Al excess in the equatorial plane of our Galaxy at the satellite observations (Mahoney et al., 1984). The results of this, probably, were the differentiation of all silicate or silicate-content bodies and heliocenric structuring the Main asteroid belt (Grimm *et al.*, 1993; Ghosh *et al.*, 2006).

We consider also the possibility of evaluation of asteroid surface composition by remote methods, in particular, by spectrophotometry. It is based, first of all, on comparison of asteroid reflectance spectra with the laboratory reflectance spectra of their probable analogues. As is known, such materials are terrestrial rocks, minerals, other compounds and available meteorites being in most cases fragments of asteroids. The samples are studied, as a rule, in a fragmented state – down to the smallest particles. Only in this way, high-quality diffuse reflectance spectra of solid substance carrying information on its composition can be obtained. It is also taken into account the well-known fact that the surfaces of terrestrial planets, their moons and asteroids are covered with a layer of rocks crushed at collisions. The method was described in details in previous publications (e. g., Adams *et al.*, 1970; Busarev, 1999; Busarev, 2011b).

Then results of the mentioned methods can be used as initial conditions in solving analytical equations in modeling or as the cornerstones in numerical simulations.

## REFLECTANCE SPECTRA OF SELECTED ASTEROIDS AND ANALOG SAMPLES

As we discuss here the nature of C-type asteroids, consider spectral characteristics of their possible analogs, the main of which are carbonaceous chondrites and terrestrial hydrated silicates (e. g., Gaffey *et al.*, 1989, 2002).

**Carbonaceous chondrites** of the most primitive types have very low albedo – of about 0.05-0.07 in the photometric band *V* (e. g., Zellner *et al.*, 1977). They are represented here by seven well known samples, reflectance spectra of which were measured in our work (Busarev *et al.*, 2002) (Table 1 and Fig. 1 (A, B). As can be seen from Fig. 1, the most primitive carbonaceous chondrites have a concave shape of reflectance spectra in the larger part of the visible range. It is explained by existence of wide one or two absorption bands centered from 0.65 to 0.95 μm because of electronic charge transfer $Fe^{2+} \rightarrow Fe^{3+}$ (Burns et al., 1972; Platonov, 1976; Burns, 1993). These absorption bands are typical for hydrated silicates abundant in the matrix of carbonaceous chondrites. It should be made an important remark that at increased content of $Fe_2O_3$ in the terrestrial pyroxenes and olivines, this band can even mask their diagnostic absorption bands at 1 micron (Adams, 1975).

**Serpentines** and other phyllosilicates are the main constituents of the matrix of carbonaceous chondrites (e. g., Dodd, 1981). So, it is important to describe their diagnostic spectral features. As discovered in our previous work (Busarev *et al.*, 2008), in addition to abovementioned absorption bands at 0.65 - 0.95 μm, reflectance spectra of low-iron (up to ab. 3 wt. % of FeO) serpentines, these are lizardite and chrysotile, demonstrate an intense absorption band at 0.44 μm (up to ~25% in relative intensity) (Fig. 1, C, Table 1). Interestingly, such type of serpentines frequently occurs among terrestrial rocks which are at an initial stage of aqueous alteration (Deer *et al.*, 1963) and, for the reason, could origin or even be widespread on rock-ice celestial bodies. It is found that equivalent width of the band has a strong correlation (up to ~90%) with the content of $Fe^{3+}$ (octahedral and tetrahedral) in corresponding samples (Busarev *et al.*, 2008). Then we concluded that the absorption band could be used as an indicator of $Fe^{3+}$ and the serpentines themselves.

Thus, described absorption bands of phyllosilicates centered at 0.44 μm and 0.65 - 0.95 μm may be useful means in carrying out spectral investigations of C-type and other primitive asteroids in the most usable visible and near-infrared range (0.4–1.1 μm). As for the absorption band of $Fe^{3+}$ at 0.44 μm, it is not seen in reflectance spectra of carbonaceous chondrites (see Figs. 1, A, B) due to, probably, their very low albedo. However, it could be observed in reflectance spectra of C-type and similar bodies if the matter is in a mixture with more light materials. Additional indicators of oxidized substances can be weak absorption bands centered at 0.60 and 0.67 micron discovered in reflectance spectra of S-type asteroids and oxidized Fe-, Fe-Ni–compounds and minerals of spinel group, representing complex oxides of Fe, Mg, Al, Cr (Hiroi *et al.*, 1996). Of great importance for investigations of rock-ice bodies is an absorption band of water ice and/or bound water in reflectance spectra at 3.0 µm (e. g., Lebofsky et al., 1981). In particular, measurements in the spectral range managed to find for the first time water ice on the surface of the Galilean satellites of Jupiter (Moroz, 1965), detect atypical hydrated silicates on the surface of a significant number of asteroids of high-temperature types (Rivkin et al., 1995, 2000; 2002).

**Reflectance spectra of selected asteroids** of the Main belt were obtained at different phases of rotation of the bodies (Figs. 2-7; see also Table 2) and represented here from a longer list in the previous publication (Busarev, 2011b) as the most spectacular examples. These spectra were measured from August 2003 to April 2010 at 1.25-m telescope and a CCD-spectrograph (with CCD-camera SBIG ST-6) of the Crimean Observatory of the SAI in the range of 0.40-0.92 micron with a spectral resolution of ~8Å. Registration of each spectrum of an object in the operating range, as a rule, carried out serially in two intervals of wavelengths (of 0.40-0.67 and 0.65-0.92 microns, or vice versa) and took from 10-15 minutes to a half of hour depending on the exposition time. Initial processing of observational data is carried out by a standard way with the use of a software package shipped with CCD-camera SBIG ST-6. Binding and calculation of the wavelength scale in the spectral range were carried out on the Balmer lines of hydrogen in the spectrum of the primary standard α Lyr. Subsequent operations connected with calculation of reflectance spectra of asteroids, their smoothing and other corrections, carried out with the help of standard programs "Excel" and "Origin".

Together with the asteroids, standard stable stars were observed. The stars were used for definition of the function of spectral transparency of terrestrial atmosphere on each observation night. These stars were selected to be at the same time analogs of the Sun on the spectrophotometric parameters (16 Cyg B, HD 117176 and HD10307) (Hardorp, 1980; Cayrel de Strobel, 1996). The average time (universal time), the conditions of observations of asteroids and standard stars (the difference of air masses), the ephemerides of the asteroids, as well as the relative mean standard deviations in reflectance spectra are given in Table 2. The calculated reflectance spectra were smoothed out (by the method of the «running box averaging ») over the entire spectral range to eliminate a high frequency noise component (Busarev, 2011b). In the cases of large errors in asteroid reflectance spectra near the borders of the working spectral range (within ~500 Å), as well as at residual telluric bands, a polynomial approximation performed. And, finally, the reflectance spectra were normalized to unit at 0.55 µm (if they were obtained in the whole spectral range of 0.40-0.92 micron or only in the short-wave part). If the reflectance spectrum was obtained only in the long-wave part of the spectrum, it was normalized at $\lambda = 0.65$ µm. In order not to complicate the figures, the errors of measurements for each of the spectra are given only in Table 2 (the last three columns). On the average, as can be seen, the relative errors in the central part of the visible range of the reflectance spectra are about 1-2% and rise up to 5-7% near the borders of the used spectral range.

**(1) Ceres** is the largest asteroid ($T_{rot.}= 9.^h074$; $p_v = 0.113$; D = 848.4 km; Sp = G, C, C) and one of dwarf planets (since 2006) according to the classification of IAU (here and further in brackets, next parameters of the asteroid are given: the rotation period ($T_{rot.}$) (Batrakov *et al.*, 2000), the IRAS geometric albedo ($p_v$) and diameter (D) (Tedesco *et al.*, 2004) and taxonomic classifications (Sp) according to the three most popular systems developed by Tholen (Tholen, 1984), Bus (Bus, 1999) and Bus-DeMeo (DeMeo *et al.*, 2009) in chronologic order). Thus, the presently adopted taxonomic or spectral classification of Ceres is "C". It means that its spectral features are similar to those of carbonaceous chondrites (e. g., Gaffey et al., 1989). Observations of Ceres with the Hubble Space Telescope (HST) have adjusted its average geometric albedo (0.090) (Li et al., 2006) and diameter (974,6 x 909,4 km) (Thomas et al., 2005). The average density of Ceres was determined as 2.2±0.1 gcm$^{-3}$ (Carry et al., 2008) which is close to that of carbonaceous chondrites.

Obtained reflectance spectra of Ceres show (Fig. 2 A, B) that its surface substance has a different composition. The reflectance spectra 1 and 2 (Fig. 2 A) and 3-5 (Fig. 2 B) correspond to opposite sides of the asteroid (see Table 2). Some parts of Ceres's surface (spectra 1 and 2) are typical for asteroids of G-type. The other part of it is like to CBF-type asteroids of (spectra 3–5), that is, may have a lower-temperature (carbonaceous-chondritic) composition. A significant change of the shape of the spectra – from convex (curves 1 and 2) to concave (curves 3-5) testifies about the change in the degree of hydration (or oxidation) of the substance, affecting the intensity of the $Fe^{2+} - Fe^{3+}$ charge-transfer absorption bands in the visible range. Such variations may be the result of various heating of the asteroid surface during its early thermal evolution and/or the subsequent major shock events. It is confirmed by the significant variations in the albedo of Ceres discovered during photometric observations of its surface with the HST (Li et al., 2006). Thus in average, a low albedo and reflectance spectra characterize Ceres as a body with predominantly carbonaceous-chondritic composition.

One more important issue in a common general interest to Ceres, as a dwarf planet among asteroids or a possible survived asteroid parent body (APB), is the next. Observed shape and surface roughness of Ceres point to a central mass concentration indicative of differentiation (Thomas *et al.*, 2005). Taking into account this fact and a possible initial carbonaceous-chondritic or rock-icy composition of the dwarf planet, models (e. g., Castillo-Rogez *et al.*, 2010) show its differentiated structure consisting of the central silicate or hydrosilicate core, water ice mantle and icy crust. However, as follows from observational data, we see that carbonaceous-chondritic materials are on the surface of Ceres. An interpretation of this contradiction will be offered in the section devoted to the hypothesis on the origin of C-type asteroids and carbonaceous chondrites. Much hope was placed on the NASA Dawn spacecraft, which must reach immediate surroundings of Ceres in 2015.

**(2) Pallas** is the second largest asteroid ($T_{rot.}$= 7.$^h$813; $p_v$ = 0.159; D = 498.1 km; Sp = B, B, B). The spectral type "B" is close to "C" (e. g., Tholen *et al.*, 1989) that means that Pallas has in average a slightly bluer reflectance spectrum then Ceres. During observations of Pallas with the HST, the value of its diameter was specified (582x556x500 km) and distinct variations of the color and albedo were discovered across the asteroid surface (Schmidt *et al.*, 2009). In spite of the similarity of the spectral characteristics of Ceres and Pallas, the latter has a higher average density, of 3.4±0.9 gcm$^{-3}$ (Carry *et al.*, 2010). This may be a sign of distinctions in the internal structure of these bodies. However, the simulation of the thermal evolution of Pallas shows that its internal structure could be formed as similar to that of Ceres, that is, the silicate core and a water ice mantle, though less thick then Ceres's one (e. g., Schmidt *et al.*, 2012). Obviously, to elucidate the question, further studies of Pallas are needed.

Obtained reflectance spectra of Pallas (Fig. 3) close in rotational phases (Table 2) show that there are also changes in composition and/or in the degree of hydration (oxidation) of the surface matter with rotation of the asteroid. The spectra 1 and 2 are similar to that of C-type, and the spectrum 3 corresponds to F-type (Tholen *et al.*, 1989). At the same time, adopted average type of the asteroid is "B" close to "C". A convex shape of the reflectance spectra 1 and 3 testifies probably about presence of a carbonaceous-chondritic material on the surface of Pallas.

**(4) Vesta** ($T_{rot.}$= 5.$^h$342; $p_v$ = 0.423; D = 468.3 km; Sp = V, V, V) is the third largest and the brightest asteroid, as its geometric albedo is 3-4 times more then $p_v$ of exceeding it in size Ceres and Pallas. The diameter of Vesta was specified as 530 km with the HST (Thomas *et al.*, 1997). Discovering of the overall spectral similarity of Vesta and howardite-eucritediogenite (HED) basaltic achondrites (McCord et al., 1970 ) was the first and the most convincing evidence of an igneous origin and magmatic differentiation of Vesta.

The most notable features in the obtained reflectance spectra of Vesta (Fig. 4) are their predominantly convex shape typical for a solid body with a high-temperature mineralogy (Tholen *et al.*, 1989; Gaffey *et al.*, 1989) as well as a prominent absorption band of orthopyroxenes at 0.90 μm caused by spin-allowed electronic transitions in ions $Fe^{2+}$ located in the M2 positions which is confirmed by the presence of a weaker pyroxenic absorption bands at 0.51 μm, caused by spin-forbidden electronic transitions in ions $Fe^{2+}$ (Burns *et al.*, 1972; Platonov, 1976). It is in a good agreement with observational spectral data and interpretations of other authors (e.g., McCord et al., 1970; McFadden et al., 1977, Gaffey, 1997) characterized Vesta's surface content as a basaltic and mainly pyroxenic one (e. g., Gaffey, 1997). At the same time, we found some unusual spectral features in reflectance spectra of Vesta (Fig. 4 A, curves 1 and 2). These are a slightly concave shape its reflectance spectra in the range 0.40-0.73 μm reminiscent of reflectance spectra of carbonaceous chondrites and combined with the previous one a weak absorption band at 0.43-0.44 μm which could testify presence of $Fe^{3+}$ included in hydrated silicates. We suppose that the listed spectral features are evidences of an admixture and/or separate units of carbonaceous-chondritic matter in some places of the surface of Vesta. Unusual presence of hydrated silicates on the surface of Vesta (taking into account its high-temperature evolution) was discovered at ground-based observations of the asteroid in the range about 3.0 μm (Hasegawa *et al.*, 2003). Interestingly, the observations of Vesta were made at close values of sub-Earth longitudes on the asteroid as ours (Fig. 4 A, curves 1 and 2; Table 2). We have undertaken also study of Vesta with the spectral-frequency method based on a row of its rotationally resolved reflectance spectra and the frequency analysis of corresponding values of the equivalent width of an absorption band at 0.44 μm considering its as an indicator of hydrated silicates (Busarev *et al.*, 2007). Thus, it was found that there are spots of hydrated silicates on the surface of Vesta and, if adopting an average diameter of the asteroid as 550 km, more than 50% such spots have sizes from 50 km to 13 km (Prokof'eva-Mikhailovskaya *et al.*, 2008). These dimensions are similar to those of craters formed on terrestrial planets and asteroids (Ivanov et al., 1999). Significance of the result is discussed further.

NASA Dawn spacecraft obtained a detailed color map and a large amount of new information on Vesta. The spacecraft has been orbiting Vesta since July 15, 2011 until the end of August 2012. The available data on Vesta should be carefully analyzed to answer the question why the evolutionary path of Vesta strikingly differs from that of close to it in size Ceres and Pallas.

**(8) Flora** is S-type asteroid ($T_{rot.}$= 12.$^h$799; $p_v$ = 0.243; D = 135.9 km; Sp = S, Sw) and a head of a large family of dynamically and spectrally similar small bodies (Zappala *et al.*, 1994). An idea based on observational spectral results was proposed that Flora is a differentiated S-type asteroid (Gaffey, 1984). At the same time, a characteristic absorption band of hydrated silicates at 3 μm was discovered in Flora's reflectance spectrum (Eaton *et al.*, 1983). As it follows from this, Flora has contradictory spectral features.

Close in rotational phase reflectance spectra of Flora are virtually identical (Fig. 5 A, spectra 1-3, Table 2) and largely in compliance with its adopted spectral types (Tholen, 1989; DeMeo *et al.*, 2009). Spectral type "Sw" means that the reflectance spectra of the Flora have higher slope in the continuum than the conventional S-asteroids (DeMeo et al., 2009). This is confirmed only by the spectrum 4 (Fig. 5 B), obtained for other side of the asteroid (see Table 2). It may be an indication of the variation of exposure age of different parts of the asteroid surface. Olivine-pyroxenic (at 0.90 μm) absorption band and a weaker pyroxenic one (at 0.52 microns) are noticeable in all spectra and are characteristic for an S-type asteroid. At the same time, there is an absorption band at 0.45-0.46 μm ($Fe^{3+}$?) (Fig. 5 A, curves 1-3) and a distortion at 0.76-0.86 μm (Fig. 5 B) ($Fe^{2+} \rightarrow Fe^{3+}$?) which probably indicate presence on the surface of Flora of oxidized and/or hydrated compounds. In addition, there are absorption bands at 0.62 and 0.67 μm (Fig. 5, curve 4) arising probably due to presence of oxidized metal compounds on Flora.

Significance of the mentioned results growths taking into account spectral similarity and large number of Flora's family asteroids counting about 500 members (Zappala *et al.*, 1994).

**(10) Hygiea** is one of the largest C-type asteroids ($T_{rot.}$ = 27.$^h$623; $p_v$ = 0.072; D = 407.1 km; Sp = C, C, C). The first two reflectance spectra were measured close in time (about a half of hour) and at good atmospheric conditions (Fig. 6 A, curves 1 and 2). Therefore, spectral differences in these spectra are within the measurement errors (see Table 2). This is to be expected for the slowly rotating asteroid. However, the shape of the Hygiea's spectra is not consistent with the spectral type «C» with a low-temperature mineralogy (Gaffey *et al.*, 1989; Gaffey *et al.*, 2002) but rather corresponds to reflectance spectra of an asteroid of S-type (Tholen *et al.*, 1989) and the mineral olivine (e. g., Platonov, 1976). The reflectance spectra of Hygiea 3 and 4 (Fig. 6 B) registered on another date (Table 2) also have a small difference in rotation phase (Table 2). Though there are some differences (approximately 10-20%) between the spectra in the range of 0.65-0.91 μm, they mostly consistent with the spectral type «C». Low-temperature mineralogy of Hygiea is confirmed in the spectra by a wide absorption band at 0.55-0.80 μm ($Fe^{2+} \rightarrow Fe^{3+}$?) and a weak absorption band of $Fe^{3+}$ at 0.44-0.45 μm (Fig. 6 B). The last reflectance spectrum of Hygiea 5 (Fig. 6 C) corresponds to its side opposite to that represented by spectra 1 and 2 shown in Fig. 6 A. The shape of the spectrum is similar to those of asteroids B - and F-types, close to C-type (Tholen *et al.*, 1989). Thus, the reflectance spectra of Hygiea obtained at different rotational phases testify about significant heterogeneity of its surface composition, connected, probably, with its thermal evolution and/or collisional processes (Busarev, 2011a).

**(21) Lutetia** ($T_{rot}$= 8.$^h$166; $p_v$ = 0.221; D = 95.8 км; Sp = M, Xk, Xc) was a target of ESA Rosetta spacecraft. The spacecraft had the flyby of the asteroid on 10 July 2010 and made valuable measurements. According to the taxonomic classification of Tholen (Tholen, 1989), the asteroid is of M-type. It is of Xk-type (a moderate "red" continuum shorter than the 0.75 μm and a flat continuum at longer wavelengths) in the classification of Bus (Bus, 1999). At last, it is of Xc-type in the classification of Bus-ДеМео (with the continuum slope from small to the middle and a slightly convex shape of spectrum).

Previously, we have observed spectra and calculate reflectance spectra of Lutetia at different oppositions and rotational phases (e. g., Busarev, 2010) which show considerable changes. For instance, the reflectance spectra 1-3 registered on November 5-6, 2004 (Table 2) have a concave shape like that of spectra of primitive BGF-type (close to C-type) asteroids (Tholen *et al.*, 1989). As already noted, such a shape of reflectance spectrum in the visible range is typical for carbonaceous chondrites enriched with hydrated silicates as a sign of electronic charge transfer in iron cations ($Fe^{2+} \rightarrow Fe^{3+}$) (see Figs. 1 A, B and 7 A). A similar result was obtained by other authors at the same asteroid opposition (Nedelcu *et al.*, 2007). In addition, there is one more indicator of the $Fe^{3+}$ content in the surface matter of Lutetia – a weak absorption band at 0.43-0.45 μm (Fig. 7 A, B). Reflectance spectra of Lutetia characterizing its diametrically opposite side had essentially another shape with a significant positive slope on November 7-8, 2004 (Fig. 7 B). Common weak details in these spectra are absorption bands at 0.43-0.45 μm ($Fe^{3+}$) and 0.52 ($Fe^{2+}$) μm, as well as one at 0.71-0.72 μm ($Fe^{2+} \rightarrow Fe^{3+}$ ?). And, finally, on 1-2 December 2008, the reflectance spectra of (Fig. 7 C, curves 10 and 11) were similar to those of an S-type (Fig. 7 C, curve 10) and an M-type asteroids (Fig. 7 C, curve 11 (Tholen *et al.*, 1989). Such interpretation is confirmed by discovering of an absorption band of hydrated silicates at 3 μm in reflectance spectra of Lutetia (Rivkin *et al.*, 2000). We have also performed the spectral-frequency investigations of the asteroid based on its rotationally resolved reflectance spectra and a frequency analysis of corresponding values of the equivalent width of an absorption band at 0.44 μm as an indicator of hydrated silicates (Busarev *et al.*, 2007). At simplifying assumptions (about spherical shape of the asteroid and its diameter ~100 km), it was found that at the level of signal to noise ratios S/N = 4-5 dimensions of spots of hydrated silicates on the surface of the Lutetia are predominantly in the range from 10 to 3 km (e. g., Busarev *et al.*, 2007). It agrees probably with spots of a dark matter seen on an image of Lutetia's surface obtained by the ESA Rosetta spacecraft on 10 July 2010 (http://www.esa.int/images/6_Lutetia_OSIRIS_LAM_2,0.jpg).

Thus, Lutetia has controversial spectral properties. This corresponds to the changes in the average composition of Lutetia's surface substance from carbonaceous- chondritic to high-temperature mineral and/or metallic one. One should probably keep on the previously adopted taxonomic classification of Lutetia as an M-type or, at least, as a differentiated body. First of all, it takes into account its relatively high IRAS geometric albedo (0.22). Then, the presence of hydrated silicates included probably in carbonaceous chondrites on its surface is strange. It points to an external source of the material. Consider this in the next sections.

## EARLY WATER DIFFERENTIATION OF SILICATE-ICY BODIES AND THEIR FATE NEAR PROTO-JUPITER

Analytical calculations show (e. g., Busarev *et al.*, 2003) that extensive internal water oceans could form due to $^{26}Al$ decay and exist at ~4 °C for a long time (~5 Myr or even more after origin of CAIs) on large (*R*>100 km) Edgeworth-Kuiper objects and other rock-ice bodies in the early solar system outside the "snow-line". The analytical estimations agree with results of numeric simulation of the early thermal evolution of similar bodies (e. g., McKinnon *et al.*, 2008). After the early water differentiation and freezing, the objects and pre-planetary bodies originated in the formation zones of giant planets were probably sources of icy and rock-organic matter after their breaking up at collisions with asteroid parent bodies (APBs) in the main asteroid belt (MAB). As a consequence, intensive fluxes of the grinded materials might have changed considerably surface mineralogy of APBs shortly after their accretion (Busarev, 2004).

There is no doubt that a high surface density of matter in the formation zones of giant planets (especially, in Jupiter's one) led to runaway accretion of the planets themselves and smaller neighbouring bodies (e. g., Safronov, 1969; Lissauer *et al.*, 2009; Benvenuto *et al.*, 2009). This ensured a short time of growth of the large Jupiter zone bodies (JZBs) and, hence, enough $^{26}$Al in their interiors for complete melting water ice except for the ~10-km crust (Busarev *et al.*, 2003). If primordial JZBs similarly to comet nuclei consisted of "dirty" ice, sedimentation of solid particles with a density >1 g cm$^{-3}$ (silicates and heavy organics) in the water ocean was accompanied by the phyllosilicate formation (mainly due to the process of serpentinization). Thus, a result of water differentiation of JZBs could be accumulation of a considerable silicate-organic core with a size up to ~0,7$R$ (Busarev *et al.*, 2003). The exogenous reaction of serpentinization of silicates could supply additional heat and give an intense release of $H_2$ and $CH_4$ gases (Rosenberg *et al.*, 2001; Wilson *et al.*, 1999) in the water oceans of JZBs. Mutual collisions could also support a prolonged warm state of the bodies and repeated resurfacing of their crusts. It is important to note that the factors should have created heterogeneous, porous and weak structure of JZBs. Nearly at the same time, these bodies started to be ejected out of the zone (including to the MAB) at velocities 1-3 to 10-15 kms$^{-1}$ when proto-Jupiter reached a few masses of the Earth (Safronov, 1969; Safronov et al., 1991). JZBs-APBs' collisions could have different consequences depending on the relative velocity and sizes of the bodies. When large relative velocities and/or sizes of JZBs (surpassing considerably these of APBs), they sweep out the counterparts from the MAB at collisions (Safronov *et al.*, 1991; Ruskol *et al.*, 1998). On the contrary, when relative velocities of JZBs penetrating the MAB near the lowest boundary (1-3 kms$^{-1}$) corresponding to eccentricity of their orbits within 0.3-0.4 (Safronov *et al.*, 1991; Ruskol *et al.*, 1998), JZBs-APBs' collisions end with fragmentation of the firsts for their weak mechanical strength. It should be emphasized that the last type of collisions was more probable for JZBs having lower relative velocities as they had been moving in less elongated orbits and could penetrate into the MAB at a higher rate and for a longer time. Consequences of this would be survival at collisions and delivery to the MAB a huge amount of grinded icy, hydrated silicate and organic matter of JZBs (likely resembling CI-meteorites).

## ABOUT C-TYPE ASTEROIDS AND CARBONACEOUS CHONDRITES

Based on the described observational and theoretical results, a hypothesis is proposed on the origin of C-type asteroids and carbonaceous chondrites (Busarev, 2011c; Busarev, 2012). Its meaning is the following. Re-accretion of thick layers of JZBs' matter dispersed at collisions on the surfaces of a considerable part of APBs could lead to formation of the largest C-type asteroids (as well as 1 Ceres and 2 Pallas). If it is true, the largest C-type asteroids have silicate cores (either differentiated, or not, depending on APBs' accretion time which controls the quantity of trapped $^{26}$Al in the matter). Then, smaller C-type asteroids could be relict fragments of JZBs. Similarly, different groups of carbonaceous chondrites (CM2, CO, CV, etc.) including chondrules might have been formed in the same and/or subsequent processes of JZBs-APBs' and mutual APBs' collisions. The hypothesis is in accordance with a proposed mechanism of origin of chondrules in carbonaceous chondrites at collisions of asteroid-sized bodies (Urey, 1952; Hutchison *et al.*, 2005). It is also important that found values of shocks experienced the meteorites (Scott *et al.*, 1992) are in accordance with the most probable relative velocities of JZBs-APBs' collisions, that is, 1-3 kms$^{-1}$.

The hypothesis is confirmed by the next main facts: 1) a predominant and all-embracing heliocentric distribution of C-type asteroids in the MAB (Gradie *et al.*, 1982) 2) a growth of the relative number of asteroids of C-complex to the outer edge of the MAB and the orbit of Jupiter (Bus *et al.*, 2002) 3) unusual presence of hydrated silicates on the surfaces of (1) Ceres and (2) Pallas if the bodies are regarded to be differentiated according to the recent observational data (e. g., Thomas et al., 2005; Carry *et al.*, 2010); 4) discoveries of atypical hydrated silicates on the surfaces of many asteroids of high-temperature types (or having predominantly high-temperature mineralogy), M-, S-, E- and V-types (Rivkin *et al.*, 2000; Busarev, 2002, 2010, 2011b), 5) internal structure of carbonaceous chondrites (i. e., the absolute predominance of phyllosilicates in their matrix, sizes of silicate particles in the matrix close to those of IDPs, etc.) (Dodd, 1981; Weisberg *et al.*, 2006); 6) residual unidirectional magnetization of carbonaceous chondrites (e. g., Stacey *et al.*, 1961; Buttler, 1972). In terms of the proposed hypothesis, the last could be explained by re-accretion of carbonaceous-chondritic matter originated in JZBs on the surfaces of early APBs having melted interiors and, hence, magnetic fields.

Nevertheless, all these issues deserve further investigations.

## CONCLUSIONS

Thus, a hypothesis based on observational and theoretical results about the origin of C-type asteroids and carbonaceous chondrites is put forward. Asteroids of C-type and close BGF-types could form from hydrated silicate-organic matter accumulated in the cores of water-differentiated bodies existed in the growth zone of Jupiter and, possibly, Saturn. The gravitational scattering of such bodies by Jupiter at its final stage of formation to the main asteroid belt might have led to fragmentation and re-accretion of their primitive matter on the surfaces of many asteroids and/or asteroid parent bodies. Similarly, asteroids of other primitive types (D and P) characterized, probably, elevated content of organics could origin from the matter of small bodies accreted and evolved in the growth zones of Uranus and Neptune and then dispersed by the giant planets to the asteroid belt. The hypothesis makes clear a row of long-standing puzzling facts, the main of which are as follows. The low-albedo and carbonaceous-chondritic surface properties of (1) Ceres contradict to its probable differentiated structure and icy crust (e. g., Thomas *et al.*, 2005; Castillo-Rogez *et al.*, 2010), but it could be explained by the process of primitive matter fall. Atypical hydrated silicates (probably, as a component of carbonaceous-chondritic matter) are found on the surfaces of many asteroids of high-temperature types (Rivkin *et al.*, 1995; Rivkin *et al.*, 2000; Busarev, 1998; Busarev, 2002) that may be a consequence of prolonged precipitation of carbonaceous-chondritic matter on their surfaces. Some carbonaceous chondrites are unidirectionly magnetized (e. g., Stacey *et al.*, 1961; Butler, 1972) that may be an indication of their stay on the surface of a differentiated body having strong magnetic field.

Hydrodynamical numeric simulations show that embryos of Jupiter and Saturn could experience inward migration (because of exchange of their angular momentums to that of a massive gas-disk of the early solar system) and then a rapid outward migration and mutual orbital separation on a time scale of no more than 5 My after CAIs' formation (Morbidelli *et al.*, 2007; Pierens *et al.*, 2008; Morbidelli A. *et al.*, 2010; Walsh *et al.*, 2011). For uncertainty of the gas-disk parameters, the heliocentric distances of Jupiter at the beginning and end of the inward migration are unknown. The values may be from 3-5 (Pierens *et al.*, 2008) to ~1.5 AU (Morbidelli A. *et al.*, 2010). For instance, the small mass of Mars and early repopulation of the MAB from two very different parent populations of planetesimals (from Mars' and Jupiter's zones) could be explained (Morbidelli A. *et al.*, 2010; Walsh *et al.*, 2011). Regardless of whether or not this very early stage of the MAB repopulations was, the proposed hypothesis describes a possible mechanism of primitive matter delivery to the MAB from neighboring formation zones, especially from Jupiter's one. It represents an alternative way of origin of heliocentric structure of the MAB having partly overlapping distributions of asteroid spectral types.

## ACKNOWLEDGMENTS

The work was partly supported by the Russian Foundation for Basic Research (project no. 12-02-90444 -Ukr).

Table 1. Samples of investigated carbonaceous chondrites and terrestrial phyllosilicates (according to Busarev *et al.*, 2002; Busarev *et al.*, 2008).

| Names and types of the samples | The sample numbers | Physical state and particle sizes of the samples | |
|---|---|---|---|
| **Carbonaceous chondrites:** | | | |
| Orguel (CI) | 2476 | powdered | <0.25 mm |
| Mighei (CM2) | 1856 | powdered | <0.20 mm |
| Murchison (CM2) | 15044 | powdered | <0.25 mm |
| Boriskino (CM2) | 198 | powdered | <0.25 mm |
| Kainsaz (CO3) | 15265 | powdered | <0.25 mm |
| Allende (CV3) | 15037 | powdered | <0.25 mm |
| Grosnaja (CV3) | 73 | powdered | <0.30 mm |
| **Phyllosilicates:** | | | |
| Serpentine (100%) | 1a | powdered | <0.15 mm |
| Serpentine (63%) | 4a | powdered | <0.15 mm |
| Serpentine (not measured) | 4b | powdered | <0.15 mm |
| Serpentine (99%) | 10a | powdered | <0.15 mm |
| Serpentine (94%) | 28 | powdered | <0.15 mm |
| Brucite (67%) & Serpentine (33%) | 30 | powdered | <0.15 mm |
| Serpentine (94%) | 2540 | powdered | <0.15 mm |

Footnotes: Meteorite groups of carbonaceous chondrites are given in brackets; names of mineral samples are given according to their main constituents determined by X-ray measurements (in brackets) (Busarev *et al.,* 2008); content of serpentine in the sample 4b was not measured because of its insufficient quantity.

Table 2. Observational parameters of selected asteroids and solar analog stars (according to Busarev, 2011b).

| Names of objects and numbers of asteroid reflectance spectra | Date | UT (h m s) | α (h m s) | δ (° ' ") | Δ (AU) | r (AU) | φ (°) | V (m) | ω, L | M(z) | σ$_1$ | σ$_2$ | σ$_3$ |
|---|---|---|---|---|---|---|---|---|---|---|---|---|---|
| 16 Cyg B | 2007 10 04 | 21 03 04 | 19 41 52 | +50 31 00 | - | - | - | 6.2 | - | 1.394 | - | - | - |
| **(1) Ceres** ($T_{rot}$ = 9.$^h$07417) | | | | | | | | | | | | | |
| 1 | 2007 10 05 | 02 38 31 | 03 33 19 | +09 04 06 | 2.035 | 2.843 | 14.1 | 8.0 | 0.000 | 1.436 | 0.023 | 0.014 | 0.023 |
| 2 | 2007 10 05 | 02 52 29 | 03 33 19 | +09 04 05 | 2.035 | 2.843 | 14.1 | 8.0 | 0.026 | 1.491 | 0.017 | 0.008 | 0.018 |
| 3 | 2009 04 02 | 20 44 45 | 10 34 53 | +25 43 57 | 1.743 | 2.552 | 16.0 | 7.5 | 0.449 | 1.094 | 0.013 | 0.005 | 0.012 |
| 4 | 2009 04 02 | 22 27 46 | 10 34 51 | +25 43 47 | 1.743 | 2.552 | 16.0 | 7.5 | 0.638 | 1.303 | 0.021 | 0.007 | 0.021 |
| 5 | 2009 04 03 | 00 08 40 | 10 34 49 | +25 43 36 | 1.744 | 2.552 | 16.0 | 7.5 | 0.824 | 1.867 | - | - | 0.012 |
| HD 117176 | 2009 04 03 | 00 49 02 | 13 28 26 | +13 46 44 | - | - | - | 5.0 | - | 1.379 | - | - | - |
| **(2) Pallas** ($T_{rot}$ = 7.$^h$8132) | | | | | | | | | | | | | |
| 1 | 2007 10 03 | 20 33 59 | 22 07 35 | -02 20 23 | 2.353 | 3.202 | 11.1 | 9.2 | 0.000 | 1.595 | 0.033 | 0.023 | 0.042 |
| 2 | 2007 10 03 | 20 53 28 | 22 07 34 | -02 20 33 | 2.353 | 3.202 | 11.1 | 9.2 | 0.042 | 1.658 | 0.026 | 0.010 | 0.030 |
| 3 | 2007 10 03 | 21 09 03 | 22 07 34 | -02 20 41 | 2.353 | 3.202 | 11.1 | 9.2 | 0.075 | 1.724 | 0.031 | 0.011 | 0.028 |
| 16 Cyg B | 2007 10 03 | 21 20 01 | 19 41 52 | +50 31 00 | - | - | - | 6.2 | - | 1.469 | - | - | - |
| 16 Cyg B | 2008 10 28 | 17 59 30 | 19 41 52 | +50 31 00 | - | - | - | 6.2 | - | 1.310 | - | - | - |
| **(4) Vesta** ($T_{rot}$ = 5.$^h$342) | | | | | | | | | | | | | |
| 1 | 2008 10 28 | 23 24 05 | 02 33 22 | +03 38 35 | 1.539 | 2.521 | 4.3 | 6.4 | 0.000; 203.9 | 1.676 | 0.036 | 0.011 | 0.023 |
| 2 | 2008 10 29 | 00 06 36 | 02 33 20 | +03 38 28 | 1.539 | 2.521 | 4.3 | 6.4 | 0.133; 251.6 | 1.984 | 0.037 | 0.010 | 0.024 |
| 5 | 2008 10 30 | 21 35 12 | 02 31 26 | +03 31 33 | 1.540 | 2.522 | 4.2 | 6.4 | 0.646; 76.2 | 1.369 | 0.017 | 0.010 | 0.019 |
| 6 | 2008 10 30 | 23 11 02 | 02 31 22 | +03 31 18 | 1.540 | 2.522 | 4.2 | 6.4 | 0.945; 183.9 | 1.666 | 0.012 | 0.010 | 0.020 |
| 7 | 2008 10 31 | 00 24 19 | 02 31 19 | +03 31 08 | 1.540 | 2.522 | 4.2 | 6.4 | 0.173; 266.2 | 1.302 | 0.033 | 0.016 | 0.024 |
| 16 Cyg B | 2008 10 30 | 18 35 30 | 19 41 52 | +50 31 00 | - | - | - | 6.2 | - | 1.460 | - | - | - |
| HD 117176 | 2005 04 03 | 23 58 15 | 13 28 26 | +13 46 44 | - | - | - | 5.0 | - | 1.249 | - | - | - |
| **(8) Flora** ($T_{rot}$ = 12.$^h$799) | | | | | | | | | | | | | |
| 1 | 2005 04 03 | 20 18 01 | 07 40 06 | +24 50 07 | 1.847 | 2.227 | 26.4 | 10.6 | 0.000 | 1.510 | 0.030 | 0.020 | 0.015 |
| 2 | 2005 04 03 | 20 24 29 | 07 40 07 | +24 50 06 | 1.847 | 2.227 | 26.4 | 10.6 | 0.008 | 1.547 | 0.024 | 0.017 | 0.020 |
| 3 | 2005 04 03 | 20 30 02 | 07 40 07 | +24 50 06 | 1.847 | 2.227 | 26.4 | 10.6 | 0.016 | 1.579 | 0.025 | 0.023 | 0.020 |
| 4 | 2009 04 05 | 01 20 01 | 14 16 07 | -03 35 40 | 1.575 | 2.538 | 7.8 | 10.0 | 0.857 | 1.825 | 0.031 | 0.025 | 0.032 |
| HD 117176 | 2009 04 05 | 00 40 33 | 13 28 26 | +13 46 44 | - | - | - | 5.0 | - | 1.368 | - | - | - |
| 16 Cyg B | 2007 10 04 | 21 03 04 | 19 41 52 | +50 31 00 | - | - | - | 6.2 | - | 1.394 | - | - | - |
| **(10) Hygiea** ($T_{rot}$ = 27.$^h$623) | | | | | | | | | | | | | |
| 1 | 2007 10 04 | 22 42 11 | 00 27 01 | +08 51 24 | 2.370 | 3.367 | 1.7 | 10.2 | 0.000 | 1.307 | 0.046 | 0.012 | 0.047 |
| 2 | 2007 10 04 | 23 16 58 | 00 27 00 | +08 51 17 | 2.370 | 3.367 | 1.7 | 10.2 | 0.021 | 1.390 | 0.094 | 0.019 | 0.072 |
| 3 | 2008 11 26 | 02 09 35 | 04 44 36 | +25 14 24 | 2.516 | 3.494 | 2.6 | 10.4 | 0.340 | 1.628 | 0.024 | 0.006 | 0.100 |
| 4 | 2008 11 26 | 03 11 02 | 04 44 34 | +25 14 19 | 2.516 | 3.494 | 2.6 | 10.4 | 0.377 | 2.170 | 0.023 | 0.008 | 0.065 |
| 5 | 2008 12 02 | 02 13 08 | 04 39 24 | +25 01 25 | 2.507 | 3.492 | 0.9 | 10.3 | 0.556 | 1.937 | 0.080 | 0.018 | 0.049 |
| HD 10307 | 2008 11 25 | 23 25 00 | 01 41 47 | +42 36 48 | - | - | - | 4.9 | - | 1.432 | - | - | - |
| HD 10307 | 2008 12 01 | 22 59 15 | 01 41 47 | +42 36 48 | - | - | - | 4.9 | - | 1.422 | - | - | - |
| 16 Cyg B | 2004 11 05 | 17 55 08 | 19 41 52 | +50 31 00 | - | - | - | 6.2 | - | 1.221 | - | - | - |
| 16 Cyg B | 2004 11 07 | 17 50 05 | 19 41 52 | +50 31 00 | - | - | - | 6.2 | - | 1.223 | - | - | - |
| **(21) Lutetia** ($T_{rot}$ = 8.$^h$1655) | | | | | | | | | | | | | |
| 1 | 2004 11 05 | 23 48 27 | 02 32 55 | +11 20 50 | 1.257 | 2.245 | 2.5 | 9.9 | 0.000 | 1.450 | 0.018 | 0.005 | 0.011 |
| 2 | 2004 11 05 | 23 54 26 | 02 32 54 | +11 20 50 | 1.257 | 2.245 | 2.5 | 9.9 | 0.012 | 1.488 | 0.018 | 0.003 | 0.014 |
| 3 | 2004 11 06 | 00 00 41 | 02 32 54 | +11 20 49 | 1.257 | 2.245 | 2.5 | 9.9 | 0.025 | 1.526 | 0.020 | 0.005 | 0.011 |
| 4 | 2004 11 07 | 21 07 13 | 02 31 03 | +11 15 06 | 1.263 | 2.249 | 3.3 | 10.0 | 0.545 | 1.300 | 0.037 | 0.011 | 0.016 |
| 5 | 2004 11 07 | 21 13 25 | 02 31 02 | +11 15 05 | 1.263 | 2.249 | 3.3 | 10.0 | 0.558 | 1.323 | 0.017 | 0.005 | 0.026 |
| 6 | 2004 11 07 | 21 20 54 | 02 31 02 | +11 15 05 | 1.263 | 2.249 | 3.3 | 10.0 | 0.573 | 2.062 | 0.013 | 0.007 | 0.016 |
| 7 | 2004 11 07 | 21 26 57 | 02 31 02 | +11 15 04 | 1.263 | 2.249 | 3.3 | 10.0 | 0.585 | 2.167 | 0.013 | 0.006 | 0.017 |
| 10 | 2008 12 02 | 00 00 57 | 04 26 00 | +20 35 32 | 1.440 | 2.425 | 0.9 | 10.2 | 0.133 | 1.306 | 0.026 | 0.010 | 0.046 |
| 11 | 2008 12 02 | 01 41 34 | 04 25 56 | +20 35 27 | 1.440 | 2.425 | 0.9 | 10.2 | 0.338 | 1.857 | 0.067 | 0.011 | 0.060 |
| HD 10307 | 2008 12 01 | 22 59 15 | 01 41 47 | +42 36 48 | - | - | - | 4.9 | - | 1.422 | - | - | - |

<u>Footnotes</u>: The ordinal numbers of asteroid reflectance spectra preserved from the previous publication (Busarev, 2011b); designations: UT is the universal time; α is the direct ascent of the asteroid; δ is the declination of the asteroid; Δ is the geocentric distance of the asteroid; r is the heliocentric distance of the asteroid; φ – the light phase angle of the asteroid; V – the apparent magnitude of the asteroid (the asteroid parameters taken from the MPC (IAU) site; the asteroid rotation periods taken from «Ephemerids of small planets for 2001» (Batrakov *et al.*, 2000); ω is the relative phase of the asteroid's rotation calculated on the basis of its period of rotation and the value of the Julian day at the time of observation (it was taken that ω=0 for the first spectrum of each asteroid); L – the longitude of sub-Earth point on (4) Vesta according to formula in (Cochran, Vilas, 1998); M(z) is the air mass; the relative mean standard deviation in the asteroid reflectance spectrum at corresponding wavelengths: σ$_1$ – at 0.44-0.45 μm, σ$_2$ – at 0.59-0.60 μm, σ$_3$ – at 0.84-0.85 μm.

# Figures

Busarev V. V.
Fig. 1 (A, B, C).

A

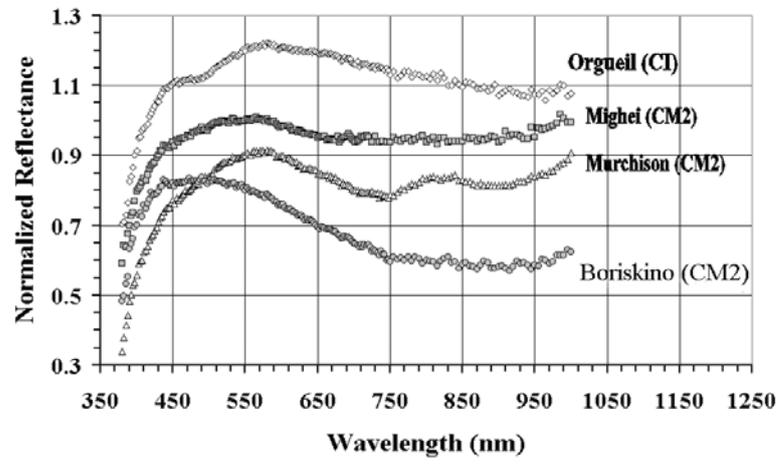

B

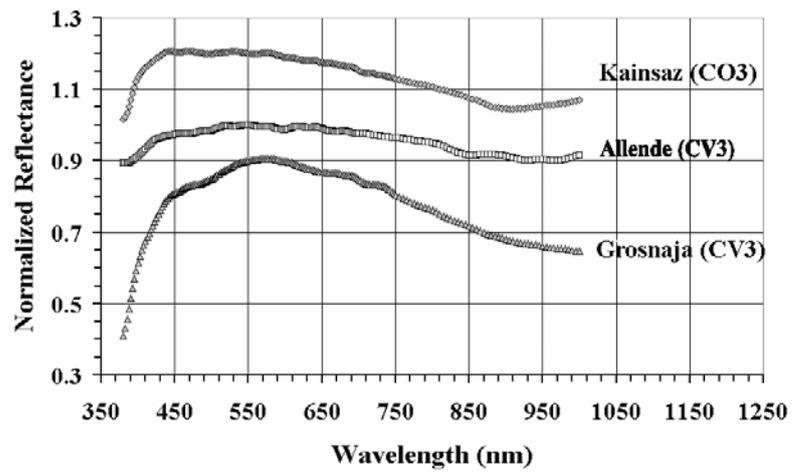

C

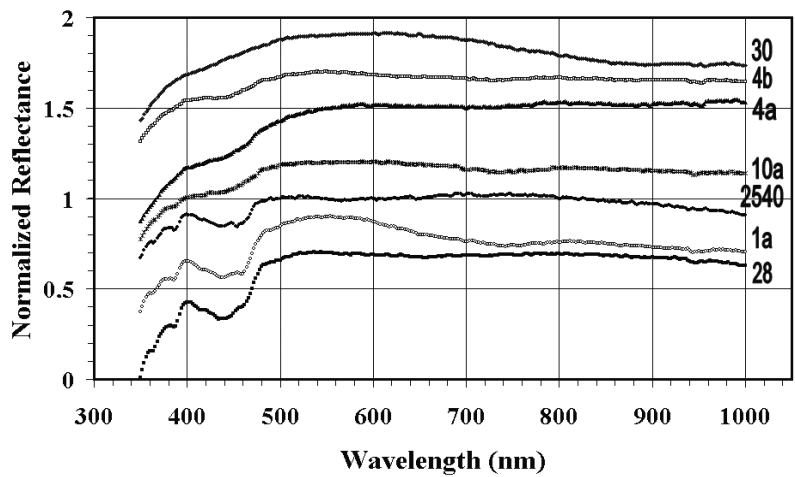

Busarev V. V.
Fig. 2 (A, B).

A

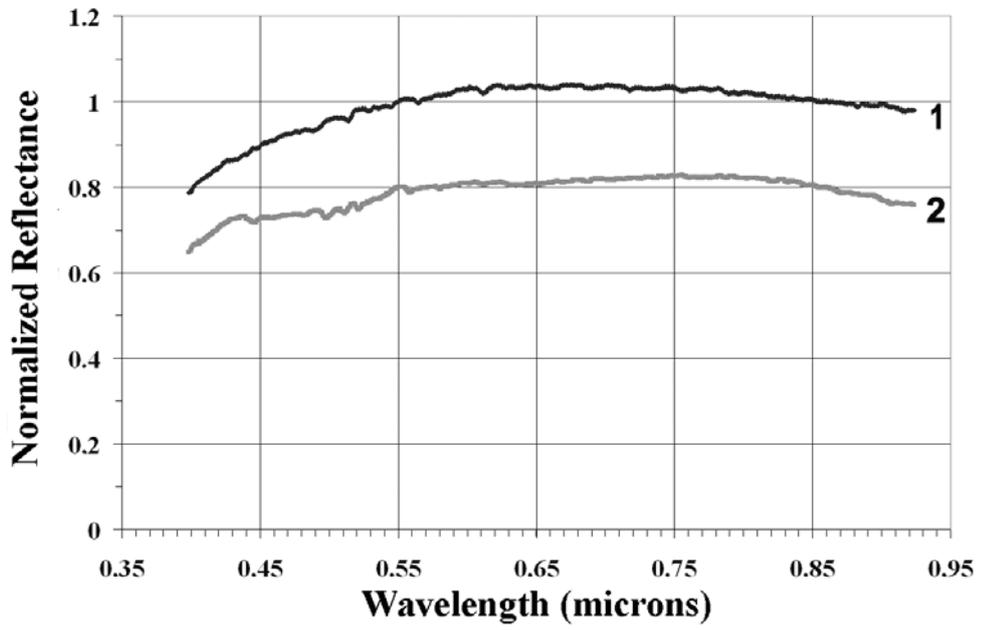

B

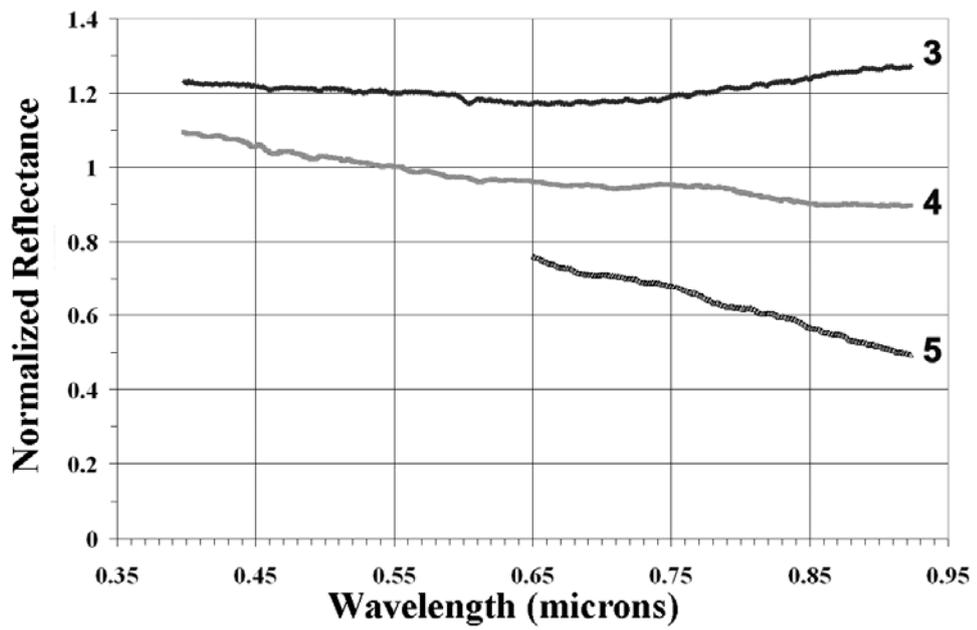

Busarev V. V.
Fig. 3.

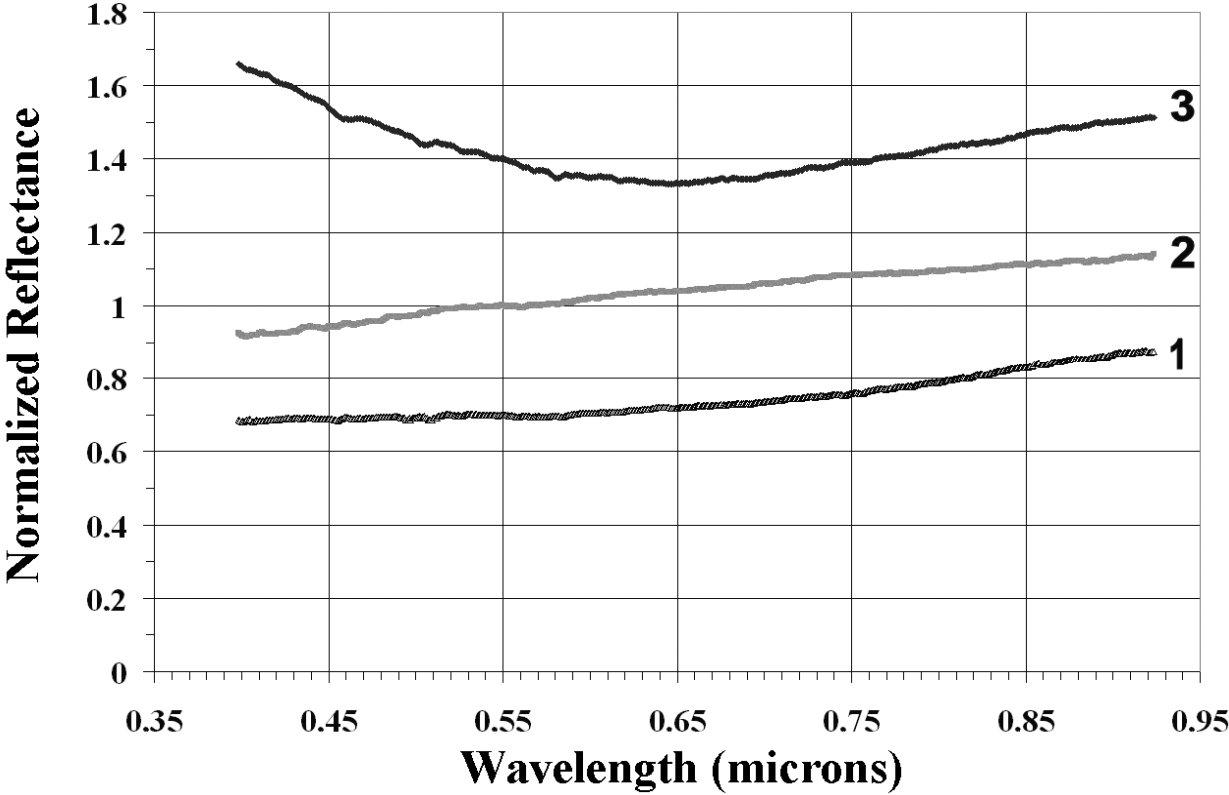

Busarev V. V.
Fig. 4 (A, B).

A

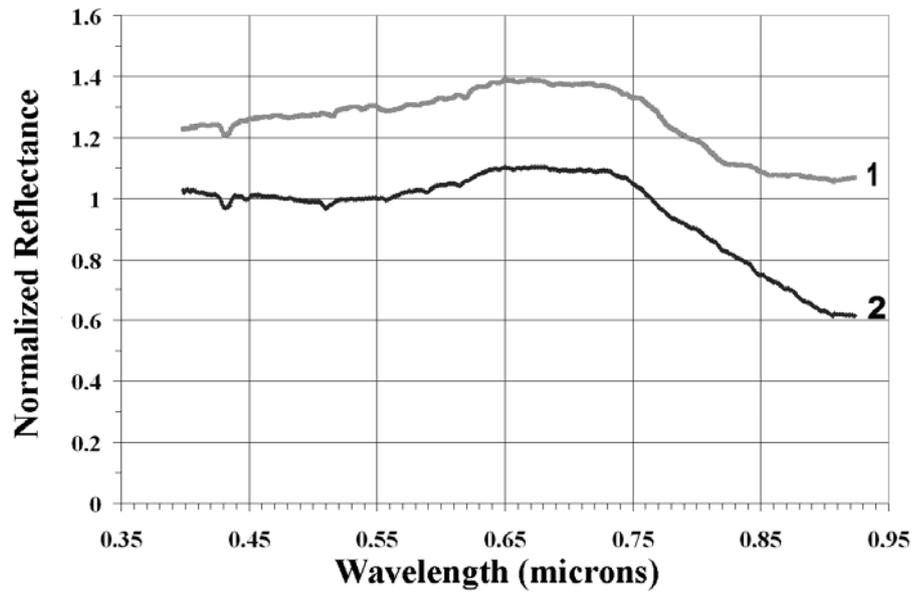

B

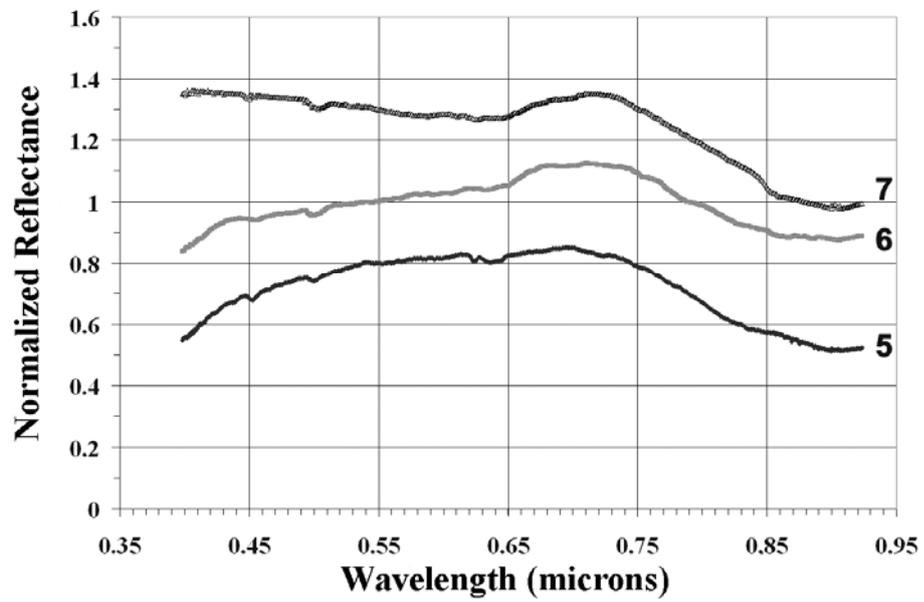

Busarev V. V.
Fig. 5 (A, B).

A

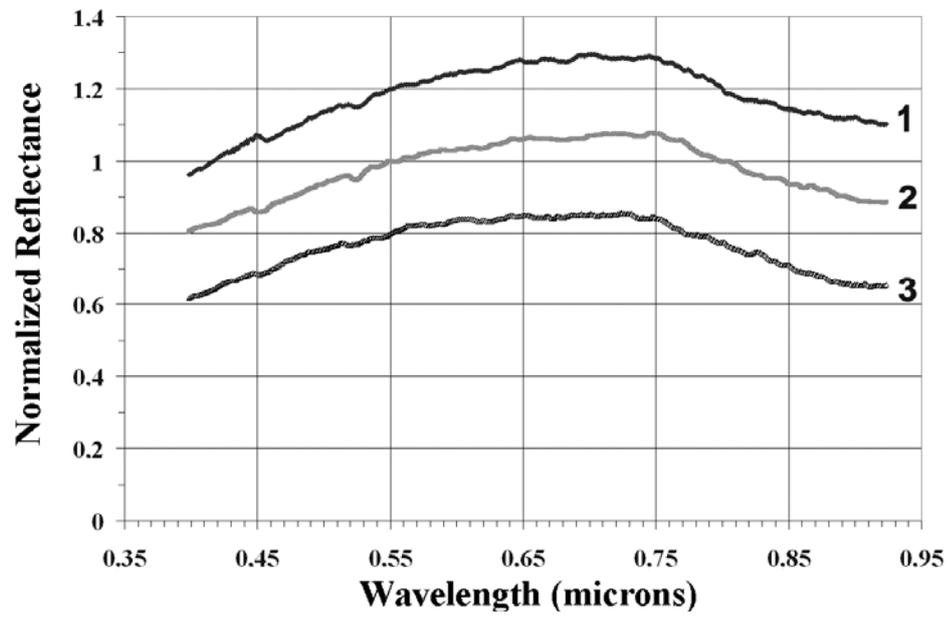

B

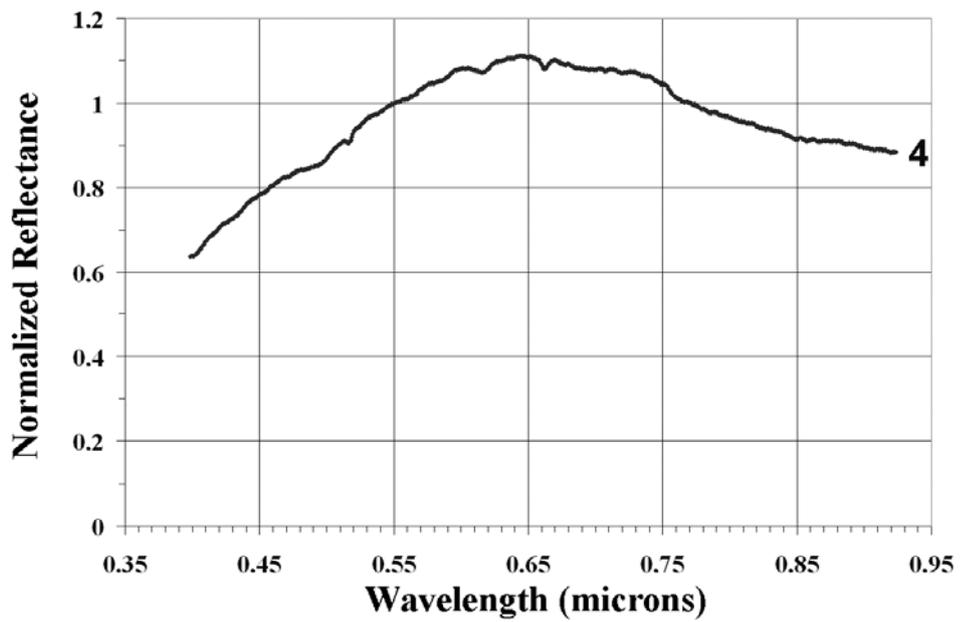

Busarev V. V.
Fig. 6 (A, B, C).

A

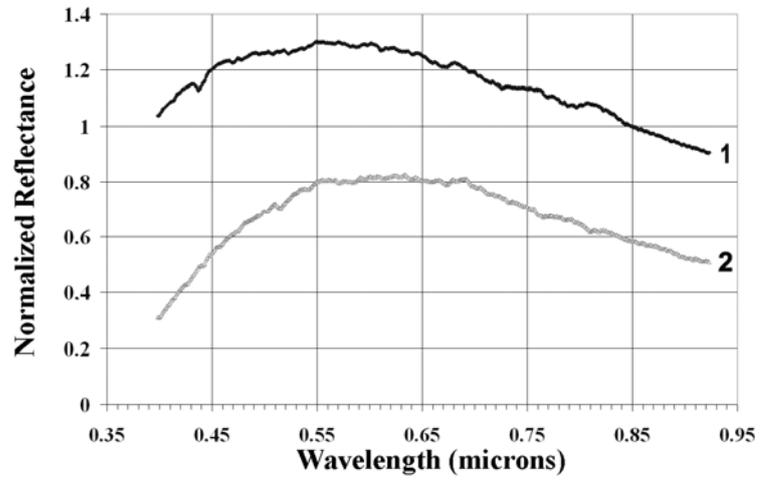

B

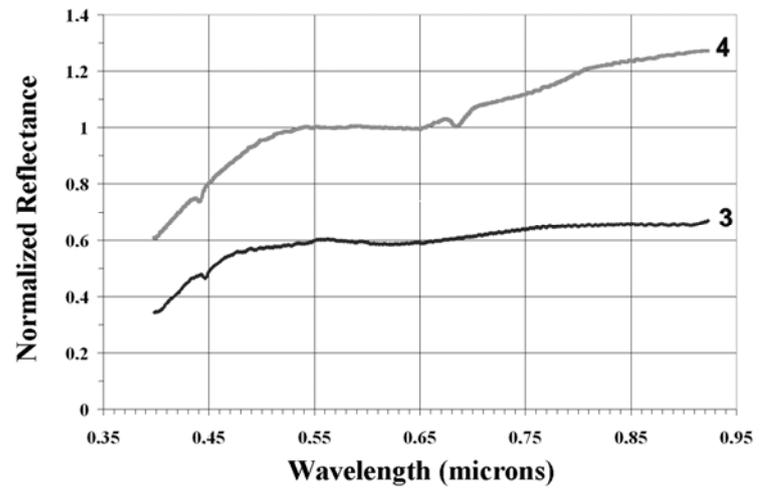

C

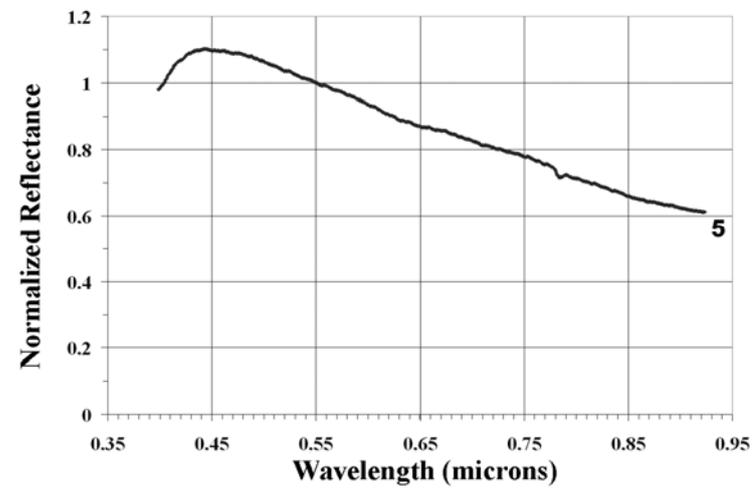

Busarev V. V.
Fig. 7 (A, B, C).

A

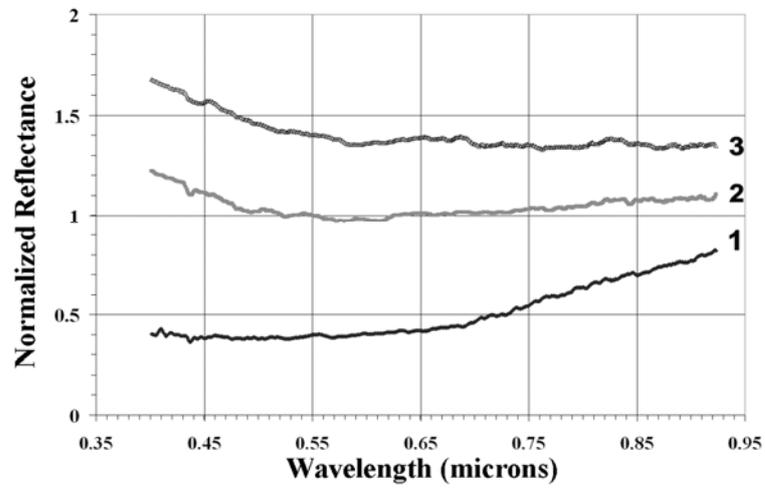

B

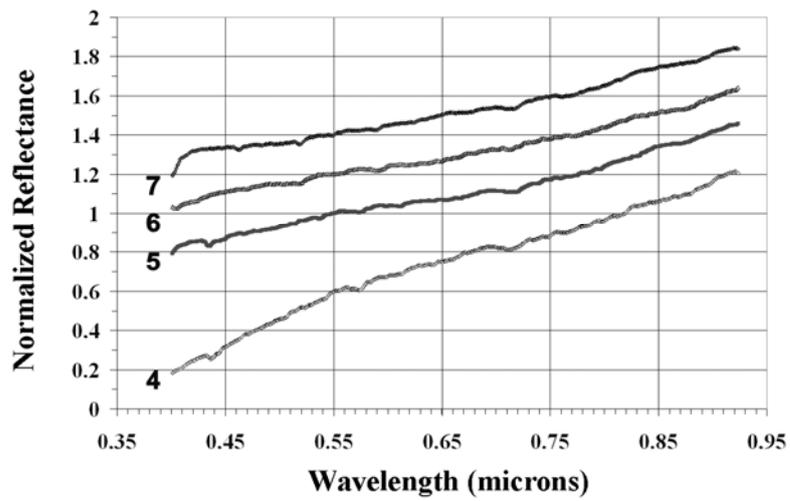

C

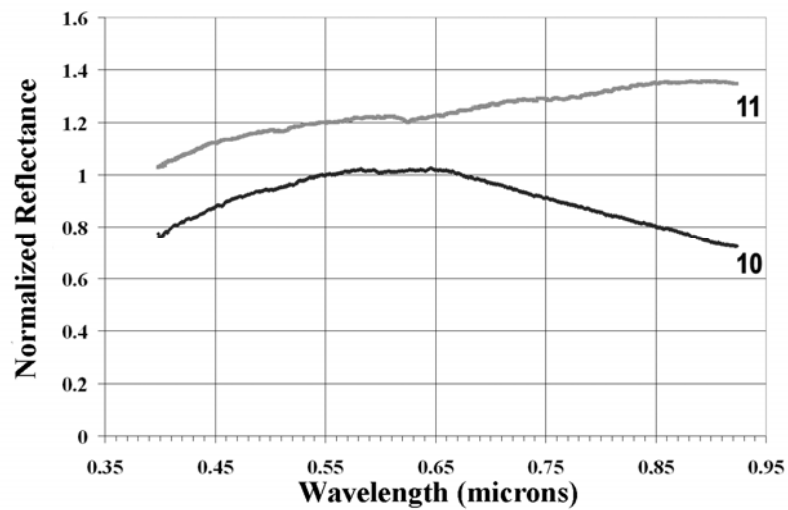

**Figure captions**

Fig. 1. (A, B, C) Normalized reflectance spectra (at 550 nm) of powdered samples of the carbonaceous chondrites Orguel (N2476) [offset by +0.2], Mighei (N1856), Murchison (N15044) [offset by -0.1], and Boriskino (N198) [offset by -0.2]. (B) Normalized reflectance spectra (at 550 nm) of powdered samples of the carbonaceous chondrites Kainsaz (N15265) [offset by +0.2], Allende (N15037), and Grosnaja (N73) [offset by -0.1]. (According to Busarev et al., 2002). (C) Normalized to unity (at 550 nm) reflectance spectra of powdered samples of terrestrial serpentines placed in order of intensity of the absorption band at 440 nm (from the top down) (according to Busarev et al., 2008).

Fig. 2 (A, B). (A) Smoothed, normalized (at 0.5503 μm) and displaced along the vertical axis (for convenience of comparison) reflectance spectra of (1) Ceres (curves 1 and 2) obtained on October 5, 2007. (A) Smoothed, normalized (at 0.5503 μm, curves 3 and 4, or at 0.6504 μm, curve 5) and displaced along the vertical axis reflectance spectra of (1) Ceres (curves 3 – 5) obtained on April 2-3, 2009. Observational parameters of the asteroid and the values of the relative medium-squares deviations (RMSD) of the spectra are given in Table 2. (According to Busarev, 2011b).

Fig. 3. Smoothed, normalized (at 0.5503 μm) and displaced along the vertical axis reflectance spectra of (2) Pallas (1 – 3) obtained on October 3, 2007. Observational parameters of the asteroid and RMSD of the spectra are given in Table 2 (according to Busarev, 2011b).

Fig. 4 (A, B). (A) Smoothed, normalized (at 0.5503 μm) and displaced along the vertical axis (for convenience of comparison) reflectance spectra of (4) Vesta (1 and 2) obtained on October 28-29, 2008. (B) Smoothed, normalized (at 0.5503 μm) and displaced along the vertical axis reflectance spectra of (4) Vesta (5 – 7) obtained on October 30-31, 2008. Observational parameters of the asteroid and RMSD of the spectra are given in Table 2. (According to Busarev, 2010).

Fig. 5 (A, B). (A) Smoothed, normalized (at 0.5503 μm) and displaced along the vertical axis reflectance spectra of (8) Flora (1 – 3) obtained on April 3, 2005. (B) Smoothed and normalized (at 0.5503 μm) reflectance spectrum of (8) Flora (curve 4) obtained on April 5, 2005. Observational parameters of the asteroid and RMSD of the spectra are given in Table 2. (According to Busarev, 2011b).

Fig. 6 (A, B, C). (A) Smoothed, normalized (at 0.5503 μm) and displaced along the vertical axis reflectance spectra of (10) Hygiea (1 and 2) obtained on October 4, 2007. (B) Smoothed, normalized (at 0.5503 μm) and displaced along the vertical axis reflectance spectra of (10) Hygiea (3 and 4) obtained on November 26, 2008. Smoothed and normalized (at 0.5503 μm) reflectance spectrum of (10) Hygiea (curve 5) obtained on December 2, 2008. Observational parameters of the asteroid and RMSD of the spectra are given in Table 2. (According to Busarev, 2011a).

Fig. 7 (A, B, C). (A) Smoothed, normalized (at 0.5503 μm) and displaced along the vertical axis reflectance spectra of (21) Lutetia (1 – 3) obtained on November 5-6, 2004. (B) Smoothed, normalized (at 0.5503 μm) and displaced along the vertical axis reflectance spectra of (21) Lutetia (4 – 7) obtained on November 7, 2004. (C) Smoothed, normalized (at 0.5503 μm) and displaced along the vertical axis reflectance spectra of (21) Lutetia (10 and 11) obtained on December 2, 2008. Observational parameters of the asteroid and RMSD of the spectra are given in Table 2. (According to Busarev, 2010).